\documentclass[aps,pra,twocolumn,groupedaddress,showpacs]{revtex4}

\usepackage{graphicx}
\usepackage{bm}
\def\expect#1{{\left\langle #1\right\rangle}}

\begin{document}

%Title of paper
\title{Four-Wave Mixing in Ultracold Atoms using Intermediate Rydberg States}

\author{E. Brekke}
%\email[]{Your e-mail address}
%\homepage[]{Your web page}
%\thanks{}
%\altaffiliation{}
%\affiliation{University of Wisconsin-Madison}
\author{J. O. Day}
%\affiliation{University of Wisconsin-Madison}
\author{T. G. Walker}
\email[]{tgwalker@wisc.edu}
\affiliation{Department of Physics, University of Wisconsin-Madison, Madison, WI 53706}

%Collaboration name if desired (requires use of superscriptaddress
%option in \documentclass). \noaffiliation is required (may also be
%used with the \author command).
%\collaboration can be followed by \email, \homepage, \thanks as well.
%\collaboration{}
%\noaffiliation

\date{\today}

\begin{abstract}
Ultracold Rb atoms were used to demonstrate non-degenerate four-wave mixing through a Rydberg state.  Continuous 5S-5P-nD two-photon excitation to the Rydberg state was combined with an nD-6P tuned laser in a phase matched geometry.  The angular dependence, spatial profile, and dependence on detuning were investigated, showing good agreement with theory.  Under optimum conditions 50 percent of the  radiation was emitted into the phase-matched direction.

\end{abstract}

% insert suggested PACS numbers in braces on next line
\maketitle

% body of paper here - Use proper section commands
% References should be done using the \cite, \ref, and \label commands
%\section{}
% Put \label in argument of \section for cross-referencing
%\section{\label{}}
%\subsection{}
%\subsubsection{}

The concept of using the strong dipole-dipole interactions between Rydberg atoms to
blockade coherent excitation of single atom qubits \cite{Jaksch00,saffman05b} has recently  been experimentally realized \cite{Urban08}.  Experimental signatures of the extension of the blockade concept to mesoscopic atomic clouds \cite{Lukin01} include observations of  suppressed excitation \cite{Tong04,Singer04,Heidemann07} and  evidence for quantum critical behavior\cite{Heidemann08,Weimer08}. Coherent coupling between ground and Rydberg atoms, an essential capability for exploitation of blockade concepts, has only recently begun to be realized\cite{Johnson08,Heidemann08,Reetz08,Weatherill08}.

One intriguing consequence of mesoscopic blockade is the collective emission of single photons, that when properly phase matched are predicted to be emitted into a diffraction-limited solid angle.  This phenomena could be used
 to produce a directional single-photon-on-demand source \cite{Saffman02} or for fast quantum-state detection or transmission \cite{Saffman05}.  Toward this end, we recently demonstrated state-selective stimulated emission detection of Rydberg atoms \cite{Day08} and showed that it can be used to study population dynamics of ultracold Rydberg clouds.  The use of light for ultracold Rydberg detection has also been accomplished using electromagnetically induced transparency \cite{Weatherill08} and shown to be very sensitive to atom-atom interactions.

In this paper we present the generation of coherent light beams using non-degenerate four-wave mixing through intermediate Rydberg states in an ultracold atomic gas.
In a low density gas we observe as much as 25\% of the emitted light isolated in a diffraction-limited solid angle under conditions of resonant Rydberg excitation.    The non-collinear excitation geometry used here is naturally compatible with producing the spatially localized excitations needed to realize collective coherent single-photon emission.  When coupled with higher-density atom clouds \cite{Sebby05} and exciting states with stronger dipole-dipole interactions \cite{Walker08}, this experiment should enter the collective single-photon regime.

\begin{figure}
\includegraphics[width=3.0 in]{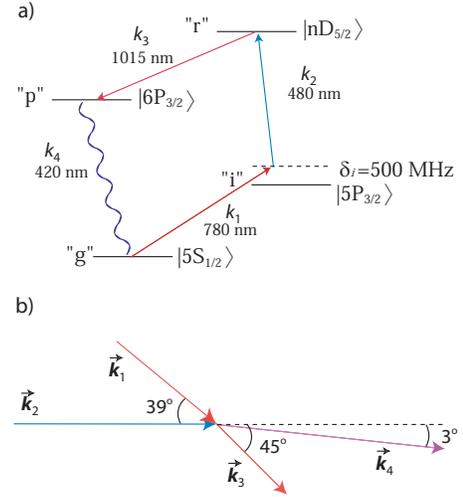}
\caption{(Color online) a) Energy levels for four wave mixing.
The $k_1$ and $k_2$ photons produce two-photon excitation to an
nD Rydberg level. The $k_3$ laser couples the Rydberg level to the 6P state. Atoms in the 6P
radiate coherently to the ground state to complete the four-wave mixing process. b) Phase-matching geometry.} \label{fig:elevel}
\end{figure}

The four-wave mixing process occurs within a $^{87}$Rb MOT, with density
$\sim$10$^{10}$ cm$^{-3}$, as described in Ref. \cite{Day08}.  The first photon ($\bm k_1$) comes from a 15 mW laser
 locked  500 MHz above the
 5S$_{1/2}$-5P$_{3/2}$ F=2 to F$^\prime$=3 transition at 780 nm.  5 mW of 480 nm light from an a frequency
doubled, amplified 960 nm ECDL is locked to a external Fabry Perot cavity, and
used as the second step, $\bm k_2$, of the excitation process and takes the atom to (or near) an nD$_{5/2}$ Rydberg state.  The third photon, $\bm k_3$,
comes from a cavity-locked 75 mW 1015 nm  laser, which brings the atoms from the nD$_{5/2}$ Rydberg
state to the 6P$_{3/2}$ state.  This results in decay photons from
the 6P$_{3/2}$ state to the ground state at 420 nm.  The relevant
energy levels can be seen in Fig. \ref{fig:elevel}a.

The directions of the three laser beams are chosen to satisfy phase-matching conditions, as shown in Fig. \ref{fig:elevel}b. In principle, a wide variety of possible geometries can be chosen, but we selected this particular one  to  give a large angle between $\bm k_2$ and $\bm k_3$, allowing for a small spatial overlap volume if desired. (A 90$^\circ$ angle is possible, giving the smallest possible overlap volume, but is incompatible with our vacuum chamber.) This results in a
phase-matched direction for the exiting 420 nm photons ($\bm k_4$) which is 3$^{\circ}$ from $\bm k_2$.  The direction of $\bm k_4$ is very insensitive to the direction of $\bm k_1$. Two
Hamamatsu H7360-01 photon counters are used to detect these 420 nm
photons, with a series of filters placed to eliminate all other
wavelengths of light from the detectors.  The detection efficiency of the filters and photodetectors is 3\%. One detector is placed
along an arbitrary or ``off-axis" direction, while a second is placed along the anticipated
phase-matched or ``on-axis" direction.
%A diagram displaying the beam geometry is shown in figure \ref{fig:geometry}.
The waists of the three beams were $(w_1,w_2,w_3)=(4.0 , 0.58 , 0.83)$ mm,
 comparable to the MOT cloud size of 0.74 mm, so that most of the MOT atoms participate in the four-wave mixing process.

%\begin{figure}
%\includegraphics[width=2.5 in]{geometry-new.eps}
%\caption{(Color online) The experimental setup, showing the phase-matching geometry of the entering laser beams, and the
%location of ``off-axis" counter 2 along an arbitrary axis, and ``on-axis" counter 1 on the predicted preferential axis. I %DON'T THINK WE NEED THIS}
%\label{fig:geometry}
%\end{figure}

Phase-matched emission was initially seen for Rydberg principal quantum number $n=28$ by moving $\bm k_1$ far off the phase-matched direction and confirming nearly equal $k_4$ detection rates in the on- and off-axis directions.  Then as $\bm k_1$ was rotated in to the phase-matched direction a factor of 10 jump in the count rate on the on-axis detector was observed, with  no change in the off-axis count rate. The width of
acceptance for phase matching was found to be $\delta\theta_1=1.6$ mrad, slightly larger than the expected 1.2 mrad. An iris placed in front of the phase matched detector reduced the solid angle of the on-axis detector by a factor of 11 while changing the count rate by only  9\%.

  When exciting to a different
n-level Rydberg state, different frequencies are involved, so the
phase-matched geometry will vary slightly.  In fact, a change of
only a few n-levels at fixed laser beam angles is enough to move the phase-matching condition outside the angular bandwidth.  After
changing from 28D$_{5/2}$ to 58D$_{5/2}$, $\bm k_1$ was
rotated until phase matching was again attained, and the rotation
angle was found to be 4.4$\pm$0.3 mrad.  This is consistent with the theoretical value of 4.1 mrad.

We scanned a razor blade across the phase-matched output
 and recorded count rate as a function of position.  The
waist of the phase-matched light was found to be 0.39 mm, slightly  (15\%) smaller than the expected diffraction limit.

Neglecting propagation effects, in the absence of blockade we can consider the four-wave mixing process to result from the spatial phasing of the atomic dipole moments as a result of the three driving fields:  $\expect{\bm d_i} = \expect{\bm d_0(\bm r_i)}\exp[i({\bm k}_1+{\bm k}_2-\bm k_3)\cdot {\bm r}_i]$, where ${\bm d}_i$ is the dipole moment of the $i$th atom and $d_0$ its magnitude as a function of atomic position $\bm r_i$. The fields from the atomic antennas constructively interfere in the far field to produce an electric field a large distance $R$ away of
\begin{equation}
E(\phi) = \frac{n\expect{d_0}k_4^2e^{ik_4R}}{R}\left(\frac{\pi
w^2}{2}\right)^{3/2}e^{-\frac{\pi^2w^2}{\lambda^2}\phi^2}
\label{eqn:dipfield}
\end{equation}
We have assumed that the dipole moment is oriented perpendicular to the plane of the lasers, and that the effective spatial distribution of the dipoles is Gaussian with standard deviation $w/2$ and peak density $n$.  The angle $\phi$ is measured with respect to the phase-matched direction.

The ratio of on-axis to off-axis powers is obtained by integrating the intensity deduced from Eq.~\ref{eqn:dipfield} over the angular distribution, and comparing to the spontaneously radiated power $P_o$ from the 6P state:
\begin{equation}
{P_{pm}\over P_o}={ {N^2_e2c\expect{ d_{0}}^2k_4^2/w^2\over  N_e\hbar c k_4\Gamma_p\rho_p}\propto{N_e|\sigma_{gp}|^2\over w^2 \rho_p}
},
\label{eqn:pmpower}
\end{equation}
where $N_e = n(\pi w^2/2)^{3/2}$ is the effective  number of participating atoms, $\rho_p$ is the fraction of atoms in the 6P state, and $\sigma_{gp}$ is the 6P-5S optical coherence density matrix element.

\begin{figure}
\includegraphics[width=3.0 in]{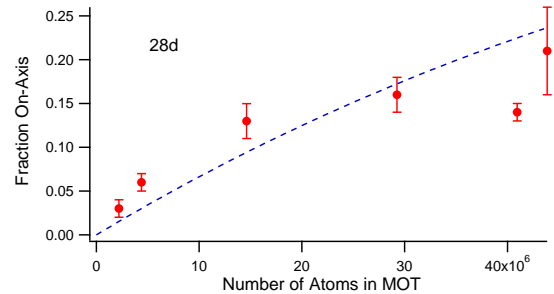}
\caption{(Color online) The fraction of light emitted in the on-axis  direction increases with the number of atoms in the MOT, as expected from the nonlinear character of four wave mixing.  The dashed line shows the model prediction.} \label{fig:percentN}
\end{figure}

From Eq.~\ref{eqn:pmpower} we expect the  phase-matched fraction to be dependent on the number of atoms, since the phase-matched power is a nonlinear process.  To verify this,  the percentage of light in the phase-matched direction was measured as a function of the number of atoms in the MOT, with the results shown in Fig. \ref{fig:percentN}.  The expected nonlinear response of the four wave mixing can be seen as a linear increase in the phase-matched fraction with number of atoms at small atom numbers.  As the cloud becomes optically thick, the phase matched light is partially scattered  due to the linear susceptibility of the atoms, and this scattered component, up to 24\% for our conditions, is observed as additional off-axis light.

For a quantitative comparison with our observations, we have developed an effective 3-level density matrix model of the four-wave mixing process.  The 5S state ``g" and the Rydberg state ``r" are coupled by an effective two-photon Rabi frequency $\Omega_1\Omega_2/2\Delta$ ($\sim$ 4 kHz) obtained by adiabatic elimination of the 5P$_{3/2}$ state.  The Rydberg state is then coupled to the 6P$_{3/2}$ state ``p" with Rabi frequency $\Omega_3\sim 1$ MHz.  In order to simulate the effects of the substantial linewidths of the lasers used (1-3 MHz), plus other possible additional broadening mechanisms, we gave the $\sigma_{rg}$  and $\sigma_{re}$ coherences effective homogeneous broadening factors of 6 and 2 MHz chosen to reproduce the observed linewidths for two-photon excitation and non-phase-matched four-wave mixing, respectively. In addition, the Rydberg state was assumed to have a shortened lifetime due to the effects of superradiance \cite{Day08}, and the 6P state was assumed to decay at its spontaneous rate. The effects of nuclear spin were accounted for by calculating effective Rabi frequencies assuming nuclear spin conservation in the Rydberg state.  Given measured intensities and atom numbers, the model then makes absolute predictions of the on- and off-axis count rates observed, and is typically within a factor of 3 of the observations.

\begin{figure}
\includegraphics[width=3.0 in]{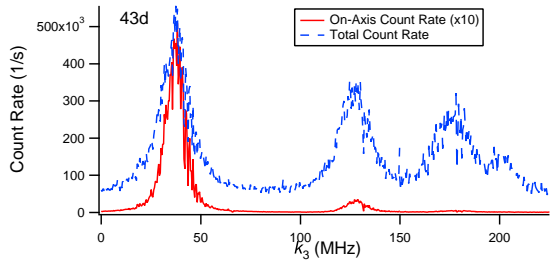}
\caption{(Color online) Count rates as a function of $k_3$, showing the hyperfine manifold of the 6P$_{3/2}$ state.  The F$^\prime$=3 state produces strikingly more on-axis light than the other F$^\prime$ levels due to its stronger coupling to the Rydberg state  and the higher branching ratio into the F=2 ground state.}
\label{fig:manifold}
\end{figure}

\begin{figure}[b]
\includegraphics[width=3.0 in]{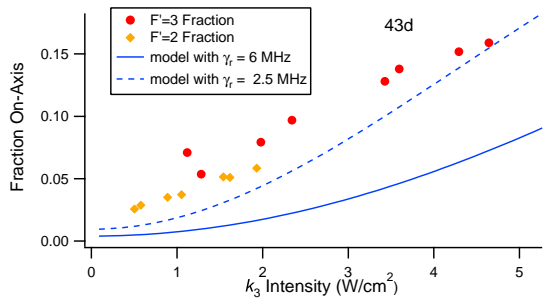}
\caption{(Color online) Fraction of on-axis light as a function of intensity, with model predictions. The intensities and on-axis fractions for F$^\prime$=2 data are scaled to reflect the reduced dipole matrix elements. The increased on-axis efficiency at higher intensity indicates the competition between  four-wave mixing and decoherence processes.  } \label{fig:powerdep}
\end{figure}

The on- and off-axis count rates as a function of $k_3$ are  shown in Fig. \ref{fig:manifold}.
 For off-axis light the F$^\prime$=3 count rate is slightly higher than the other F$^\prime$ levels, with the F$^\prime$=2 count rate being 65 percent of the F$^\prime$=3 rate. The model predicts 48\%.  The on-axis light, on the other hand, has an observed 10:1 ratio for the two states. This is partly explained by a factor of $\sqrt 2$ greater dipole matrix element for the 6P$_{3/2}$(F$^\prime$=3)$\rightarrow$5S$_{1/2}$(F=2) transition as compared to the corresponding matrix element from the F$^\prime$=2 state.  In addition there is a competition between the coherent deexcitation from the Rydberg state and the various decoherence process that occur in the Rydberg state.  This further favors the larger Rabi coupling to the F$^\prime$=3 state.
 From these effects the model
predicts a ratio of 7.7, in reasonable agreement with the observations.

Figure \ref{fig:powerdep} shows the on-axis fraction of the emitted light as  the intensity of beam $k_3$ is varied, for two different hyperfine levels of the 6P state.  The different dipole matrix elements for the two hyperfine levels are accounted for by  scaling the F$^\prime$=2 intensities by a factor of 0.48 and the on-axis fraction by 2.  The increase of the on-axis fraction with increased intensity shows that the four wave mixing process is competing with decoherence.  The model prediction is sensitive to the assumed broadening of the r-p coherence, with the data favoring a narrower (2.5 MHz) linewidth for this coherence than deduced from the off-axis spectroscopy.

\begin{figure}
\includegraphics[width=3.2 in]{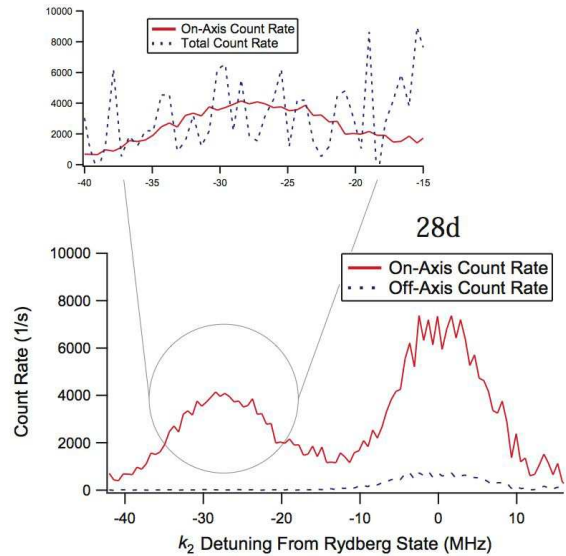}
\caption{(Color online) Count rates in on- and off-axis counters as a function of the excitation frequency, showing both the non-resonant 5S-nD excitation (left peak and inset) and resonant Rydberg state excitation with off-resonant deexcitation to the 6P state.  The $k_3$ laser was held 32 MHz above the Rydberg-6P transition. The inset scales the off-axis count rate by the solid angle to show the total amount of light emitted.
In this case the amount of on-axis light is approximately equal to the total amount of light in all other directions.} \label{fig:detscan}
\end{figure}

In order to look for another sign of decoherence, the population in the Rydberg state was reduced by detuning $k_3$ from the Rydberg-6P$_{3/2}$ resonance.  Then the $k_2$ frequency  is varied and two resonances result.  The first occurs when the excitation lasers are on resonance with the Rydberg state, giving significant Rydberg populations.  The second resonance occurs when the 3-photon process to the 6P state is on resonance, but 2-photon excitation to the Rydberg state is off resonance. An example of this data with $k_3$ tuned 32 MHz above the Rydberg-6P$_{3/2}$ F$^\prime$=3 resonance is shown in Fig. \ref{fig:detscan}. As the figure shows, the count rate on the ``off-axis" counter is
very small for off-resonant Rydberg excitation. The fraction of phase-matched light reaches 50\% of the total 6P-5S emission.  When the Rydberg states are resonantly produced, the fraction of phase matched light drops to 20\%.

\begin{figure}[t]
\includegraphics[scale=0.5]{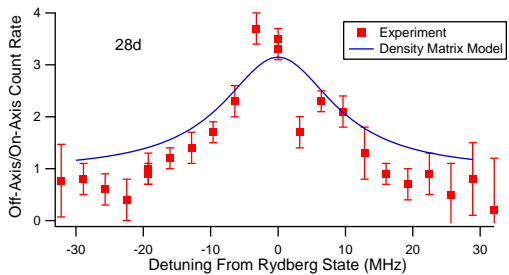}
\caption{(Color online) The ratio of off-axis counts to on-axis count rate  measured as a function of detuning from the Rydberg state.  Zero on this plot corresponds to perfect phase matching.  The efficiency of the phase-matching process is optimized when detuned from the Rydberg state.}
\label{fig:ratio}
\end{figure}

Figure \ref{fig:ratio} shows the ratio of off-axis to on-axis light as a function of 2-photon detuning, with $k_3$ adjusted to maintain 3-photon resonance.  Again, the fraction of on-axis light is about 20\% on resonance, increasing to 50\% off resonance.
  This trend is accounted for by the model as seen in the figure.

These results point out that Rydberg decoherence mechanisms are important for determining the on-axis emission efficiency.  The data presented here are all obtained under weak excitation conditions where blockade effects should be unimportant.  Under blockade conditions Rydberg-Rydberg and superradiant decoherence mechanisms are predicted to be virtually eliminated.

As a step towards using phase-matched four-wave mixing as a single photon source, we have focused the 480 nm and 1015 nm beams to 12 $\mu$m.  This results in a effective volume of $\sim$10$^{-8}$ cm$^3$, which could be blockaded for reasonable n-levels \cite{Walker08}.  When using the focused beams, phase-matching was again achieved for the various n-levels, with the same dependence on F$^\prime$ level, $k_3$ intensity, and detuning from the Rydberg state.  The maximum percentage of light achieved in the phase-matched direction was  reduced to below 1\%, as expected from Eq. \ref{eqn:pmpower}  since the relevant factor $N_e/w^2$ was reduced by a factor of 40. In the future, this could be be improved by increasing the density of the sample \cite{Sebby05}, allowing both efficient phase-matching and dipole blockade and making a single photon source possible.

\begin{acknowledgments}
This research was supported by the National Science Foundation.
\end{acknowledgments}

\bibliography{lasercooling}

\end{document}